\documentclass[showpacs,preprintnumbers,amsmath,amssymb,prb,twocolumn]{revtex4}

\def\XXint#1#2#3{{\setbox0=\hbox{$#1{#2#3}{\int}$}
     \vcenter{\hbox{$#2#3$}}\kern-.5\wd0}}

\usepackage{graphicx}
\usepackage{subfigure}
\usepackage{bm}

\begin{document}

\title{Phase competition in frustrated anisotropic antiferromagnet in strong magnetic field}

\author{O.\ I.\ Utesov$^{1,2}$}
\email{utiosov@gmail.com}
\author{A.\ V.\ Syromyatnikov$^{1}$}
\email{asyromyatnikov@yandex.ru}

\affiliation{$^1$National Research Center ``Kurchatov Institute'' B.P.\ Konstantinov Petersburg Nuclear Physics Institute, Gatchina 188300, Russia}
\affiliation{$^2$St.\ Petersburg State University, 7/9 Universitetskaya nab., St.\ Petersburg 199034,
Russia}

\date{\today}

\begin{abstract}

We discuss theoretically a frustrated Heisenberg antiferromagnet in magnetic field close to the saturation one. It is demonstrated that a small biaxial anisotropy and/or the magnetic dipolar interaction produce a delicate balance between phases with a commensurate canted, incommensurate helical (conical), and fan spin orderings. As a result, different sequences of phase transitions are realized depending on values of these small anisotropic interactions. We derive analytical expressions for critical fields and ground-state energies of the phases which are in a quantitative agreement with our and previous Monte-Carlo simulations.

\end{abstract}


\maketitle

\section{Introduction}
\label{SIntro}

Since experimental observation of the giant magnetoelectric effect the multiferroicity of spin origin has being one of the hottest topic of the contemporary condensed matter physics. \cite{Cheong2007,Tokura2009,nagaosa} The possibility arising in these materials to control electric polarization with magnetic field and magnetic ordering with electric field opens up new ways for promising technological applications (see, e.g., Refs.~\cite{Kimura2003,hur2004,mni3}).

Importantly, two of three main mechanisms of ferroelectricity of spin origin~\cite{nagaosa}, inverse Dzyaloshinskii-Moriya mechanism~\cite{Katsura2005} and spin-dependent p-d hybridization mechanism~\cite{Arima2007}, require a non-collinear magnetic ordering.
Chiral magnetic structure appears due to frustration in many multiferroics~\cite{kurumaji2019}.
The competition between different frustrating spin interactions often produces very complicated phase diagrams in the temperature-magnetic field plane which are interesting in themselves (as, e.g., in multiferroics MnWO$_4$~\cite{mnw1,Ehrenberg1997,mnw3,zh} and MnI$_2$~\cite{mni3,Utesov2017,UtesovMn}).

\begin{figure}
  \centering
  \includegraphics[width=0.9\linewidth]{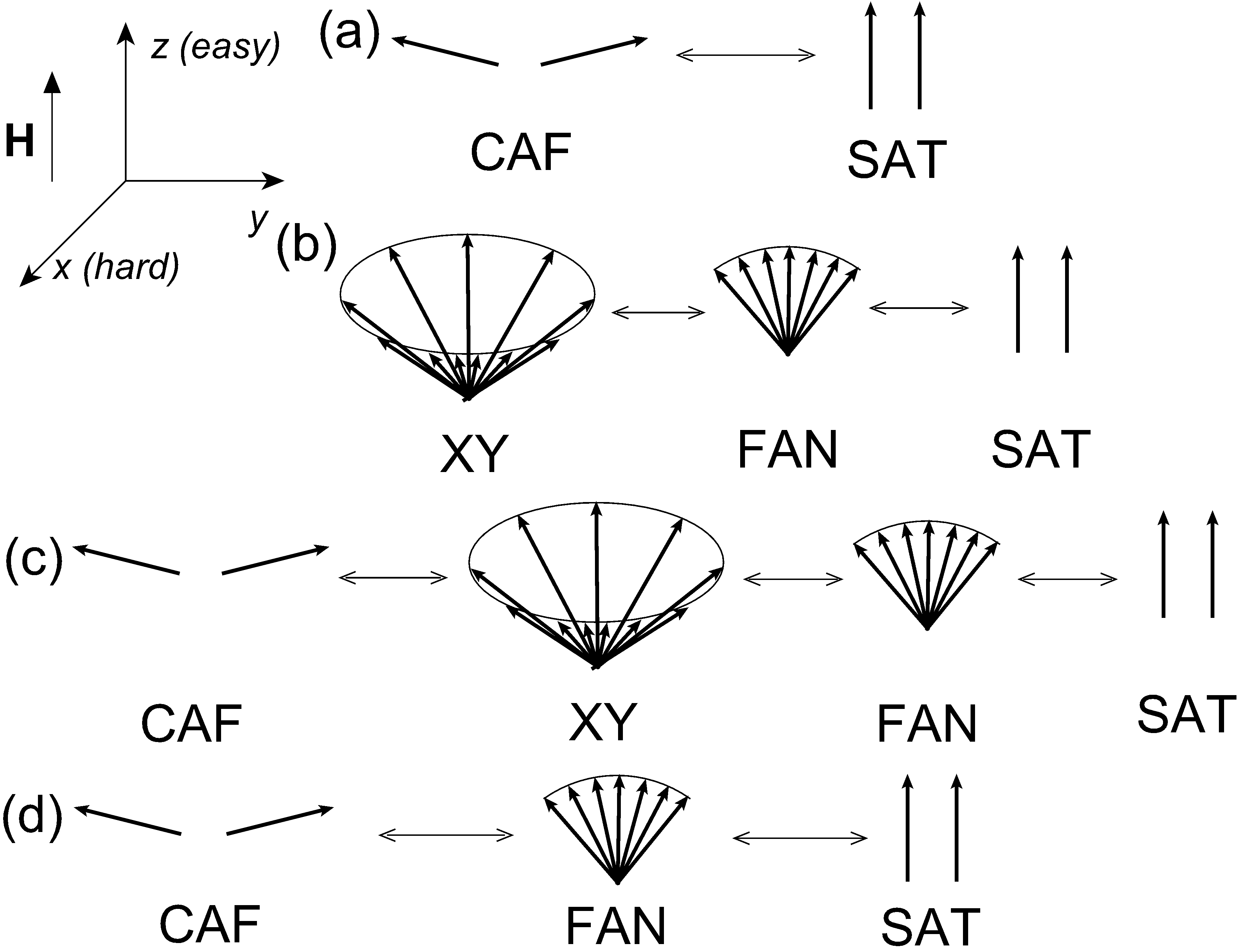}\\
  \caption{Possible sequences of phase transitions in strong magnetic field in frustrated anisotropic antiferromagnet described by the Hamiltonian~\eqref{ham1}. (a) The simplest scenario which is realized at zero anisotropy: canted two-sublattice AF phase (CAF) gradually transforms into the fully saturated state (SAT) upon the field increasing. Scenario (b) is for a weak anisotropy in the spiral plane which is perpendicular to the field direction: conical incommensurate XY phase transforms into the fan phase via an Ising phase transition; the fan state is followed by the SAT phase. Strong enough in-plane anisotropy stabilizes CAF phase at moderate fields which is followed by the XY, the fan and the SAT states (c) or by the fan and the SAT phases (d).
\label{fig1}}
\end{figure}

In the present paper, we study properties of frustrated anisotropic antiferromagnets (AFs) at small temperature in strong magnetic field close to the saturation one. The present research extends our previous discussion \cite{utesov2019} to the domain of large magnetic fields. We study a simple model of frustrated AF with a biaxial anisotropy and/or magnetic dipolar forces in the external field directed along one of the principal axes of the system. We demonstrate that small anisotropic interactions enrich the phase diagram as compared to the previous consideration \cite{Ueda2009}, in which only isotropic frustrated interactions were taken into account.
Possible sequences of phase transitions which we observe are summarized in Fig.~\ref{fig1}. We show, in particular, that an Ising-type phase transition always arises between the helical and the fan phases if there is a small axial anisotropy in the plane perpendicular to the field. We derive analytical expressions for critical fields and ground-state energies of all phases. Monte-Carlo simulations confirm our analytical findings.

The rest of the paper is organized as follows. In Sec.~\ref{STheor}, we introduce the model. Sec.~\ref{SCone} is devoted to the conical phase. In Sec.~\ref{SFan}, properties of the fan phase are addressed. We discuss in Sec.~\ref{SSeq} quite nontrivial sequences of phase transitions shown in Figs.~\ref{fig1}(c) and \ref{fig1}(d). Sec.~\ref{SComp} is devoted to comparison of our analytical results with the numerical ones of Ref.~\cite{zh} for a particular model. Dipolar forces are discussed in Sec.~\ref{SDip}. Sec.~\ref{SSum} contains our summary. In Appendix~\ref{AppendixA}, we discuss magnon spectrum in the conical phase. Appendix~\ref{AppendixB} contains some mathematical details related to the fan phase instability. In Appendix~\ref{AppendixC}, we provide some details about our Monte-Carlo simulations.

\section{Model}
\label{STheor}

We consider first a simple model of a frustrated anisotropic antiferromagnet with a small biaxial anisotropy. Dipolar interaction between spins will be taken into account in Sec~\ref{SDip}. The model Hamiltonian includes frustrated exchange interaction $\mathcal{H}_{ex}$, the anisotropy $\mathcal{H}_{an}$, and the Zeeman term $\mathcal{H}_z$
\begin{eqnarray}
 \label{ham1}
  \mathcal{H} &=& \mathcal{H}_{ex} + \mathcal{H}_{an} + \mathcal{H}_{z}, \nonumber \\
  \mathcal{H}_{ex} &=& -\frac12 \sum_{i,j} J_{ij} \left(\mathbf{S}_i \cdot \mathbf{S}_j\right), \\
  \mathcal{H}_{an} &=& - \sum_i \left[ D(S_i^z)^2 + E (S_i^y)^2\right], \nonumber \\
  \mathcal{H}_z &=& - \sum_i \left(\mathbf{h} \cdot \mathbf{S}_i\right).\nonumber
\end{eqnarray}
Here for definiteness $D > E > 0$ ($x$ and $z$ are the hard and the easy axes, respectively) and ${\bf h}=g \mu_B {\bf H}$. We assume also that there is one spin per unit cell and that the lattice is arbitrary. Performing the Fourier transform
\begin{equation}
\label{four1}
  \mathbf{S}_j = \frac{1}{\sqrt{N}} \sum_\mathbf{q} \mathbf{S}_\mathbf{q} e^{i \mathbf{q} \mathbf{R}_j},
\end{equation}
where $N$ is the number of spins in the lattice, terms in Hamiltonian \eqref{ham1} can be rewritten as follows:
\begin{eqnarray}
  \label{ex2}
  \mathcal{H}_{ex} &=& -\frac12 \sum_\mathbf{q} J_\mathbf{q} \left(\mathbf{S}_\mathbf{q} \cdot \mathbf{S}_{-\mathbf{q}}\right), \\
	\label{an21}
  \mathcal{H}_{an} &=& - \sum_\mathbf{q}\left[ D S^z_\mathbf{q} S^z_{-\mathbf{q}} + E S^y_\mathbf{q} S^y_{-\mathbf{q}}\right], \\
	\label{z21}
 \mathcal{H}_z &=& - \sqrt{N} \left(\mathbf{h} \cdot \mathbf{S}_{\bf 0}\right).
\end{eqnarray}

We assume below that the Fourier transform of exchange interaction $J_\mathbf{q}$ has two equivalent maxima at incommensurate momenta ${\bf q}=\pm {\mathbf{k}}$ so that a plane spiral arises as the classical ground state at $h=D=E=0$. The spiral plane in which spins rotate can be fixed by the small anisotropy and/or by magnetic field which can also distort the spiral order. As it was shown in Ref.~\cite{utesov2018}, at large enough magnetic fields, the conical phase is the ground state of the system with the spiral plane being perpendicular to the field direction. However a competition between the helical and the commensurate phases can appear at moderate anisotropy. \cite{utesov2019}
Let us consider first incommensurate states.

\section{Conical phase}
\label{SCone}

In this section, we describe properties of the conical phase in the external magnetic field which is directed for definiteness along $z$ axis (other field directions are discussed at the end of Sec.~\ref{SFan}).
A deformed conical helix appears in the ground state of the system at moderate magnetic fields.

For theoretical description of the cone state, we use Kaplan's helical basis (see Refs.~\cite{Kaplan1961,Maleyev2006})
\begin{eqnarray}
  \label{bas1}
  \hat{\zeta}_j &=& (\hat{a} \cos{\mathbf{k}\mathbf{R}_j} + \hat{b} \sin{\mathbf{k}\mathbf{R}_j}) \cos{\alpha} + \hat{c} \sin{\alpha},\nonumber \\
  \hat{\eta}_j &=&  - \hat{a} \sin{\mathbf{k}\mathbf{R}_j} + \hat{b} \cos{\mathbf{k}\mathbf{R}_j}, \\
  \hat{\xi}_j &=& - (\hat{a} \cos{\mathbf{k}\mathbf{R}_j} + \hat{b} \sin{\mathbf{k}\mathbf{R}_j}) \sin{\alpha} + \hat{c} \cos{\alpha},\nonumber
\end{eqnarray}
where $\hat{a}$, $\hat{b}$, and $\hat{c}$ are some mutually orthogonal unit vectors, and $\alpha$ is the cone angle ($\alpha=0$ for the plane spiral). Because $ab$ plane is the spiral plane in Eqs.~\eqref{bas1}, $\hat{a}$, $\hat{b}$, and $\hat{c}$ are directed along $x$, $y$, and $z$ axes in our system (see Fig.~\ref{fig1}), respectively. Spin at the $j$-th site is written as
\begin{equation}
\label{spin1}
  \mathbf{S}_j = S_j^\zeta \hat{\zeta}_j + S_j^\eta \hat{\eta}_j + S_j^\xi \hat{\xi}_j.
\end{equation}
Here
\begin{eqnarray}
\label{spinrep1}
  S^\zeta_j &=& S - a^\dagger_j a_j, \nonumber \\
  S^\eta_j &\simeq& \sqrt{\frac{S}{2}} \left( a_j + a^\dagger_j \right), \\
  S^\xi_j &\simeq& i \sqrt{\frac{S}{2}} \left( a^\dagger_j - a_j \right)\nonumber
\end{eqnarray}
is the Holstein-Primakoff spin representation~\cite{Holstein1940} in which square roots a replaced by unity. It is convenient to rewrite local basis vectors \eqref{bas1} as
\begin{eqnarray}
  \label{bas2}
 \hat{\zeta}_j &=& (\mathbf{A} e^{i \mathbf{k} \mathbf{R}_j} + \mathbf{A}^* e^{- i \mathbf{k} \mathbf{R}_j}) \cos{\alpha} + \hat{c} \sin{\alpha},  \nonumber \\
 \hat{\eta}_j &=& i \mathbf{A} e^{i \mathbf{k} \mathbf{R}_j} - i \mathbf{A}^* e^{- i \mathbf{k} \mathbf{R}_j}, \\
 \hat{\xi}_j &=& -(\mathbf{A} e^{i \mathbf{k} \mathbf{R}_j} + \mathbf{A}^* e^{- i \mathbf{k} \mathbf{R}_j}) \sin{\alpha} + \hat{c} \cos{\alpha}  \nonumber
\end{eqnarray}
with the use of auxiliary vectors $\mathbf{A} = (\hat{a} - i\hat{b})/2$ and $\mathbf{A}^* = (\hat{a} + i\hat{b})/2$. Then, we have from Eqs.~\eqref{spin1} and \eqref{bas2} after Fourier transform~\eqref{four1}
\begin{equation}
\label{spin2}
  \mathbf{S}_\mathbf{q} = S^A_\mathbf{q} \mathbf{A} + S^{A^*}_\mathbf{q} \mathbf{A}^* +S^c_\mathbf{q} \hat{c},
\end{equation}
where
\begin{eqnarray}
  \label{spin3}
 S^A_\mathbf{q} &=&  S^\zeta_{\mathbf{q} -\mathbf{k}} \cos{\alpha} + i S^\eta_{\mathbf{q}-\mathbf{k}} - S^\xi_{\mathbf{q}-\mathbf{k}} \sin{\alpha} ,  \nonumber\\
 S^{A^*}_\mathbf{q} &=& S^\zeta_{\mathbf{q}+\mathbf{k}} \cos{\alpha}  - i S^\eta_{\mathbf{q}+\mathbf{k}} - S^\xi_{\mathbf{q}+\mathbf{k}}  \sin{\alpha},\\
 S^c_\mathbf{q} &=&  S^\zeta_\mathbf{q} \sin{\alpha} + S^\xi_\mathbf{q} \cos{\alpha}.  \nonumber
\end{eqnarray}
Substituting Eqs.~\eqref{spin2} and \eqref{spin3} into Eqs.~\eqref{ex2}--\eqref{z21}, one obtains
\begin{eqnarray}
 \label{exch1}
  \nonumber \mathcal{H}_{ex} &=& -\frac{1}{2} \sum_\mathbf{q} \Bigl[ \left( \sin^2{\alpha} J_\mathbf{q} + \cos^2{\alpha} J_{\mathbf{q},\mathbf{k}} \right) S^\zeta_\mathbf{q} S^\zeta_\mathbf{-q}   \\
 \nonumber &+& J_{\mathbf{q},\mathbf{k}} S^\eta_\mathbf{q} S^\eta_\mathbf{-q} + \left( \cos^2{\alpha} J_\mathbf{q} + \sin^2{\alpha} J_{\mathbf{q},\mathbf{k}} \right) S^\xi_\mathbf{q} S^\xi_\mathbf{-q}  \\
  \nonumber &+& \sin{\alpha} \cos{\alpha} \left( J_\mathbf{q} - J_{\mathbf{q},\mathbf{k}} \right)  \left( S^\zeta_\mathbf{q} S^\xi_\mathbf{-q} + S^\xi_\mathbf{q} S^\zeta_\mathbf{-q} \right) \\
 \nonumber &+& i \cos{\alpha} N_{\mathbf{q},\mathbf{k}} \left( S^\eta_\mathbf{q} S^\zeta_\mathbf{-q} - S^\zeta_\mathbf{q} S^\eta_\mathbf{-q} \right) \\
 &+& i \sin{\alpha} N_{\mathbf{q},\mathbf{k}} \left( S^\xi_\mathbf{q} S^\eta_\mathbf{-q} - S^\eta_\mathbf{q} S^\xi_\mathbf{-q} \right) \Bigr],
\end{eqnarray}
where $J_{\mathbf{q},\mathbf{k}}=(J_{\mathbf{q}+\mathbf{k}}+J_{\mathbf{q}-\mathbf{k}})/2$ and $N_{\mathbf{q},\mathbf{k}}=(J_{\mathbf{q}+\mathbf{k}}-J_{\mathbf{q}-\mathbf{k}})/2$,
\begin{eqnarray}
\label{an1}
 \mathcal{H}_{an} &=& -D \sum_\mathbf{q} \left( S^c_\mathbf{q} S^{c}_{-\mathbf{q}}\right) \nonumber \\ && + \frac{E}{4} \sum_\mathbf{q} \left[ \left( S^A_\mathbf{q} - S^{A^*}_{\mathbf{q}}\right) \left( S^A_{-\mathbf{q}} - S^{A^*}_{-\mathbf{q}}\right)\right],\\
\label{z1}
  \mathcal{H}_z &=& -\sqrt{N} h S^c_\mathbf{0}.
\end{eqnarray}

Using Eqs.~\eqref{spinrep1}, we derive now a part of the bosonic Hamiltonian which contains terms with no more than two Bose operators. Notice that the biaxial anisotropy produces umklapp terms in the Hamiltonian of the type $a^\dagger_{\mathbf{q} \pm 2\mathbf{k}} a_\mathbf{q}$. We neglect these terms because they are small being of the order of $E$ and giving a contribution of higher order in $E/J$ to effects considered below (e.g., their contribution to the energy is of the third order in $E/J$). These statements are valid only if the system is not very close to the transition to the fan phase (see below). Neglecting the umklapps as well as higher order in $1/S$ terms, we obtain the following expression for the Hamiltonian~\eqref{ham1}:
\begin{equation}
\label{ham2}
  \mathcal{H} = N \varepsilon^{XY}_0 + \mathcal{H}_1 + \mathcal{H}_2.
\end{equation}
Here
\begin{equation}
\label{exy1}
  \varepsilon^{XY}_0 = - \frac{S^2}{2} \left[ \sin^2{\alpha} (J_\mathbf{0} + 2D) + \cos^2{\alpha} (J_\mathbf{k} + E)\right] -  S h \sin{\alpha},
\end{equation}
\begin{eqnarray}
\label{hamL1}
\nonumber
  \mathcal{H}_1 &=& i S\sqrt{N} \sqrt{\frac{S}{2}} (a_0 - a^+_0) \Bigl[ \sin{\alpha} \cos{\alpha} (J_\mathbf{0} - J_\mathbf{k}\\
  &&+ 2D - E) + h \cos{\alpha}/S) \Bigr] \\ \nonumber
  &&+ i E \sqrt{N} \left(\frac{S}{2}\right)^{3/2} \cos{\alpha} \Bigl[ (1+\sin{\alpha}) (a_{-2\mathbf{k}} - a^+_{-2 \mathbf{k}}) \\ \nonumber
  &&- (1-\sin{\alpha}) (a_{2\mathbf{k}} - a^+_{2 \mathbf{k}}) \Bigr],
\end{eqnarray}
and
\begin{equation}
\label{hamB1}
  \mathcal{H}_2 = \sum_\mathbf{q} \left( C_\mathbf{q} a^\dagger_\mathbf{q} a_\mathbf{q} + B_\mathbf{q} \frac{a^\dagger_\mathbf{q} a^\dagger_{-\mathbf{q}} + a_\mathbf{q} a_{-\mathbf{q}}}{2} \right),
\end{equation}
where
\begin{eqnarray}
  C_\mathbf{q} &=& \frac{S}{2} \Bigl[ 2 J_\mathbf{k} - (1 + \sin^2{\alpha}) J_{\mathbf{q},\mathbf{k}} - \cos^2{\alpha}(J_\mathbf{q} \nonumber \\
  &&+ 2D - E) - 2 \sin{\alpha} N_{\mathbf{q},\mathbf{k}} \Bigr], \\
	\label{bq}
  B_\mathbf{q} &=& \frac{S}{2} \cos^2{\alpha} (J_\mathbf{q} - J_{\mathbf{q},\mathbf{k}} + 2 D - E ).
\end{eqnarray}
Minimization of $\varepsilon^{XY}_0$ yields
\begin{equation}
\label{Ang1}
  \sin{\alpha} = \frac{h}{S(J_\mathbf{k}-J_\mathbf{0} + E - 2D)},
\end{equation}
in which case the coefficient before $a_0 - a^+_0$ in Eq.~\eqref{hamL1} becomes zero. According to Eq.~\eqref{Ang1}, the saturation field is given by
\begin{equation}
\label{ConeHs}
  h^{XY}_s = S(J_\mathbf{k}-J_\mathbf{0} + E - 2D).
\end{equation}
However we show below that the direct transition to the saturated phase from the conical one is possible only at zero anisotropy in the spiral plane, i.e., at $E=0$. At finite $E$, the requirement to dispose of terms linear in $a_{\pm2\mathbf{k}}$ and $a^+_{\pm 2 \mathbf{k}}$ in Eq.~\eqref{hamL1} leads to the stabilization of the fan phase before the transition to the fully saturated state.

We report also the absence of the symmetry $\mathbf{q} \leftrightarrow -\mathbf{q}$ at finite field in both linear terms \eqref{hamL1}, where $-2\mathbf{k}$ and $2\mathbf{k}$ momenta are not equivalent, and in the bilinear part of the Hamiltonian \eqref{hamB1}, where coefficient $C_\mathbf{q}$ is not symmetric due to the term $\propto N_{\mathbf{q},\mathbf{k}}$ (see also Ref.~\cite{zhitomirsky1996}). As it is shown below, this asymmetry plays an important role at strong fields. It is also pronounced in the magnon spectrum (see Appendix~\ref{AppendixA}).

The bare conical helix ordering described by the single momentum $\bf k$ is a subject of corrections (similar to those discussed in Ref.~\cite{utesov2018}) due to the anisotropy in the spiral plane. Terms with $a_{\pm 2\mathbf{k}}$ and $a^\dagger_{\pm 2\mathbf{k}}$ proportional to $E$ in the linear part of the Hamiltonian~\eqref{hamL1} are responsible for this effect. As in Ref.~\cite{utesov2018}, to calculate corrections we perform a shift in operators
\begin{eqnarray}
\label{shift1}
  &&a^\dagger_{2\mathbf{k}} \mapsto z_+ + a^\dagger_{2\mathbf{k}}, \quad a_{-2\mathbf{k}} \mapsto z_- + a_{-2\mathbf{k}},\nonumber\\
	&&a_{2\mathbf{k}} \mapsto z_+^* + a_{2\mathbf{k}}, \quad a^\dagger_{-2\mathbf{k}} \mapsto z_-^* + a^\dagger_{-2\mathbf{k}},
\end{eqnarray}
where $z_+$ and $z_-$ are complex numbers. In order to eliminate linear terms \eqref{hamL1} in the Hamiltonian,  $z_+$ and $z_-$ should satisfy the following set of equations:
\begin{eqnarray}
  C_{2\mathbf{k}} z_+ + B_{2\mathbf{k}} z_- = i \cos{\alpha} (1-\sin{\alpha}) E \sqrt{N} \left( \frac{S}{2} \right)^{3/2}, \\
  C_{-2\mathbf{k}} z_- + B_{2\mathbf{k}} z_+ = i \cos{\alpha} (1 + \sin{\alpha}) E \sqrt{N} \left( \frac{S}{2} \right)^{3/2} \nonumber
\end{eqnarray}
whose solution reads in the first order in $E/J$
\begin{eqnarray}
  z_+ &=& i E \sqrt{\frac{NS}{2}} \\ && \times\frac{\cos^2{\alpha}(J_\mathbf{k}- J_{2\mathbf{k}}) -  \sin{\alpha}(1-\sin{\alpha})(J_\mathbf{k}- J_{3\mathbf{k}})}{\cos{\alpha} (J_\mathbf{k}- J_{3\mathbf{k}})(J_\mathbf{k}- J_{2\mathbf{k}})}, \nonumber \\
  z_- &=& i E \sqrt{\frac{NS}{2}} \\ && \times\frac{\cos^2{\alpha}(J_\mathbf{k}- J_{2\mathbf{k}}) + \sin{\alpha}(1 + \sin{\alpha})(J_\mathbf{k}- J_{3\mathbf{k}})}{\cos{\alpha} (J_\mathbf{k}- J_{3\mathbf{k}})(J_\mathbf{k}- J_{2\mathbf{k}})} \nonumber.
\end{eqnarray}
The difference between $z_+$ and $z_-$ is due to the term arising from $N_{\pm \mathbf{2k},\mathbf{k}}$.

We derive now corrections to the spin ordering and to the ground-state energy. Bearing in mind that $\langle a^\dagger_{2\mathbf{k}}\rangle=z_+$, $\langle a_{2\mathbf{k}}\rangle = -z_+$, $\langle a^\dagger_{-2\mathbf{k}}\rangle=-z_-$, and $\langle a_{-2\mathbf{k}}\rangle = z_- $, we obtain from Eqs.~\eqref{spin3} for the spin odering
\begin{eqnarray} \label{SzCorr}
  S^z_\mathbf{q} &=& \sqrt{N} S \delta_{\mathbf{q},\mathbf{0}} \sin{\alpha} \nonumber \\ &&+ i \sqrt{\frac{S}{2}} \cos{\alpha} (\delta_{\mathbf{q},2\mathbf{k}} + \delta_{\mathbf{q},-2\mathbf{k}} ) (z_+-z_-), \\ \label{SxCorr}
  S^x_\mathbf{q} &=& \frac{\delta_{\mathbf{q},\mathbf{k}}+\delta_{\mathbf{q},-\mathbf{k}}}{2} \Bigl\{ \sqrt{N} S \cos{\alpha} \\ && + i \sqrt{\frac{S}{2}} [ (1-\sin{\alpha})z_1+(1+\sin{\alpha})z_2]\Bigr\} \nonumber \\ &&- \frac{\delta_{\mathbf{q},\mathbf{3k}}+\delta_{\mathbf{q},-3\mathbf{k}}}{2} i \sqrt{\frac{S}{2}} [(1+\sin{\alpha})z_1+(1-\sin{\alpha})z_2], \nonumber \\ \label{SyCorr}
  S^y_\mathbf{q} &=& \frac{\delta_{\mathbf{q},\mathbf{k}}-\delta_{\mathbf{q},-\mathbf{k}}}{2i} \Bigl\{ \sqrt{N} S \cos{\alpha} \\ && - i \sqrt{\frac{S}{2}} [ (1-\sin{\alpha})z_1+(1+\sin{\alpha})z_2]\Bigr\} \nonumber \\ &&- \frac{\delta_{\mathbf{q},\mathbf{3k}}-\delta_{\mathbf{q},-3\mathbf{k}}}{2i} i \sqrt{\frac{S}{2}} [(1+\sin{\alpha})z_1+(1-\sin{\alpha})z_2]. \nonumber
\end{eqnarray}
Similar to Ref.~\cite{utesov2018}, there is an elliptical distortion of the first harmonic and a small third harmonic $3\mathbf{k}$ in the spiral plane. In addition, a small variation of the cone angle appears with momentum $2\mathbf{k}$.

The correction to $\varepsilon^{XY}_0$ reads as
\begin{eqnarray}
\label{EnCorrXY1}
  \Delta \varepsilon^{XY}
	&=& \frac{\langle \mathcal{H}_1 \rangle}{2 N} \\
	&=& \frac{i E \left( \frac{S}{2} \right)^{3/2} \cos{\alpha}}{\sqrt{N}} [(1-\sin{\alpha})z_++(1+\sin{\alpha})z_-]\nonumber
\end{eqnarray}
which is of the second order in $E/J$.

As we obtain below, a subtle competition between phases arising near the saturation field when the parameter $\tilde{\alpha} = \pi/2 - \alpha$ is small being of the order of $\sqrt{E/J}$. Then, one can expand all the quantities up to the second order in $\tilde{\alpha}$. Importantly, $z_+ \sim (E/J) \tilde{\alpha} \sim (E/J)^{3/2} \ll z_- \sim (E/J)/\tilde{\alpha} \sim (E/J)^{1/2}$ in this case.
Eqs.~\eqref{SxCorr} and \eqref{SyCorr} can be rewritten as follows at $\tilde{\alpha} \ll 1$:
\begin{eqnarray}
  S^x_j &=& S \cos{\mathbf{k} \mathbf{R}_j} \left[ \tilde{\alpha} - \frac{2E}{\tilde{\alpha}(J_\mathbf{k}- J_{2\mathbf{k}})} \right],  \\
  S^y_j &=& S \sin{\mathbf{k} \mathbf{R}_j} \left[ \tilde{\alpha} + \frac{2E}{\tilde{\alpha}(J_\mathbf{k}- J_{2\mathbf{k}})} \right].
\end{eqnarray}
One can see from these equations that the elliptical distortion is reduced to a line at $\tilde{\alpha} = \tilde{\alpha}_c$, where
\begin{equation}
\label{AlCr}
  \tilde{\alpha}_c = \sqrt{ \frac{2E}{J_\mathbf{k}- J_{2\mathbf{k}}}}.
\end{equation}
This manifests a continuous transition to the fan phase. The corresponding critical field is estimated as (cf. Ref.~\cite{Nagamiya1962})
\begin{equation}
\label{HCr}
  h_{cr} = h^{XY}_s - S E \frac{J_\mathbf{k}- J_{\mathbf{0}}}{J_\mathbf{k}- J_{2\mathbf{k}}}.
\end{equation}

Notice that $z_-$ is quite large at $\tilde{\alpha} \ll 1$ that requires to take into account the neglected umklapp terms in the Hamiltonian and terms containing products of more than two Bose operators. This would complicate the consideration considerably. However, we find numerically using Monte-Carlo simulations (see Appendix~\ref{AppendixC} for some details) that Eqs.~\eqref{SzCorr}--\eqref{SyCorr} work qualitatively good in this regime.
Surprisingly, Eq.~\eqref{EnCorrXY1} reproduces even quantitatively numerical findings.
Then, we obtain below some qualitative results for the conical phase using expressions above even at fields close to the critical one. Besides, we derive in the next section an accurate expression for $h_{cr}$ (see Eq.~\eqref{FHcr1}) which differs from Eq.~\eqref{HCr} by a factor of 2 before the last term.

In Appendix~\ref{AppendixA}, we discuss the magnon spectrum and speculate on the softening of the magnon mode with momentum $-2\mathbf{k}$ which restores in the fan phase the broken $\mathbb{Z}_2$ symmetry.


We have from Eq.~\eqref{EnCorrXY1} for the energy correction at $\tilde{\alpha} \sim \tilde{\alpha}_c$
\begin{equation}
\label{EnCorrXY2}
  \Delta \varepsilon^{XY} = - \frac{S^2 E^2}{J_\mathbf{k}- J_{2\mathbf{k}}}.
\end{equation}
According to our numerics (see Appendix~\ref{AppendixC} for details), Eq.~\eqref{EnCorrXY2} works quantitatively well near $h_{cr}$ and it is used below for the analysis of the phase transition to the canted antiferromagnetic state (see Sec.~\ref{SSeq}). It should be noted that when considering the total energy of the $XY$ phase including correction \eqref{EnCorrXY2}, its minimum corresponds to the momentum slightly shifted from $\pm \mathbf{k}$. However, it can be shown that the momentum shift $\delta \mathbf{k} \sim (E/J)^2$ and the related energy variation $\sim (E/J)^4$ can be safely neglected.

\section{Fan phase}
\label{SFan}

In this section, we consider the fan phase in strong magnetic field directed along $z$ axis. The spin arrangement in the ground state is described as
\begin{eqnarray}
\label{FGS1}
  S^y_j &=& \beta S \cos{\mathbf{k} \mathbf{R}_j}, \nonumber\\
  S^z_j &=& S \sqrt{1 -\beta^2 \cos^2{\mathbf{k} \mathbf{R}_j} } =   \\
  &=& S \left[ \gamma - \varkappa \cos{2\mathbf{k}\mathbf{R}_j} - \frac{\beta^4}{64} \cos{4\mathbf{k}\mathbf{R}_j} + O(\beta^6) \right], \nonumber
\end{eqnarray}
where the small parameter $\beta \ll 1$ which vanishes in the saturated phase is introduced and
\begin{eqnarray}
  \gamma &=& 1 - \frac{\beta^2}{4} - \frac{3 \beta^4}{64}, \\
  \varkappa &=& \frac{\beta^2}{4}\left(1 + \frac{\beta^2}{4} \right).
\end{eqnarray}
Unit vectors of the local basis in the fan state~\eqref{FGS1} have the form 
\begin{eqnarray} \label{Fanbasis}
  \hat{\zeta}_j &=& \hat{y} \beta \cos{\mathbf{k} \mathbf{R}_j} + \hat{z} \left[ \gamma - \varkappa \cos{2\mathbf{k}\mathbf{R}_j} - \frac{\beta^4}{64} \cos{4\mathbf{k}\mathbf{R}_j} \right], \nonumber \\
  \hat{\eta}_j &=& -\hat{y} \left[ \gamma - \varkappa \cos{2\mathbf{k}\mathbf{R}_j} - \frac{\beta^4}{64} \cos{4\mathbf{k}\mathbf{R}_j} \right] + \hat{z} \beta \cos{\mathbf{k} \mathbf{R}_j}, \nonumber\\
  \hat{\xi}_j &=& \hat{x},
\end{eqnarray}
where we neglect $O(\beta^6)$ terms.

Then, spin components are expressed via spin components in the local basis as follows:
\begin{eqnarray} \label{Fspin1}
  S^x_\mathbf{q} &=& S^\xi_\mathbf{q}, \nonumber \\
  S^y_\mathbf{q} &=& \frac{\beta}{2} (S^\zeta_{\mathbf{q}-\mathbf{k}} + S^\zeta_{\mathbf{q} + \mathbf{k}} ) - \gamma S^\eta_\mathbf{q}  + \frac{\varkappa}{2} (S^\eta_{\mathbf{q}-2\mathbf{k}}  \nonumber \\
  &&+ S^\eta_{\mathbf{q} + 2\mathbf{k}}) + \frac{\beta^4}{128}(S^\eta_{\mathbf{q}-4\mathbf{k}} + S^\eta_{\mathbf{q} + 4 \mathbf{k}}),  \\
  S^z_\mathbf{q} &=& \frac{\beta}{2} (S^\eta_{\mathbf{q}-\mathbf{k}} + S^\eta_{\mathbf{q} + \mathbf{k}} ) + \gamma S^\zeta_\mathbf{q} \nonumber \\
  &&- \frac{\varkappa}{2} (S^\zeta_{\mathbf{q}-2\mathbf{k}} + S^\zeta_{\mathbf{q} + 2\mathbf{k}}) - \frac{\beta^4}{128}(S^\zeta_{\mathbf{q}-4\mathbf{k}} + S^\zeta_{\mathbf{q} + 4 \mathbf{k}}). \nonumber
\end{eqnarray}
Substituting Eqs.~\eqref{Fspin1} into Hamiltonian~\eqref{ham1} and using Eqs.~\eqref{spinrep1}, we obtain
\begin{equation}
\label{Fham1}
  \mathcal{H} = N \varepsilon^{FAN}_0 + \mathcal{H}_1 + \mathcal{H}_2,
\end{equation}
where
\begin{eqnarray}
\label{Fen1}
  \varepsilon^{FAN}_0 &=& - h S - \frac{S^2}{2}(J_\mathbf{0} + 2 D) \nonumber\\
	&&+ S \frac{\beta^2}{4}
  [h - S (J_\mathbf{k} - J_\mathbf{0} +2 E - 2 D) ] \nonumber\\
  && + S\frac{\beta^4}{64} [3 h + S(J_\mathbf{0} - J_{2\mathbf{k}})].
\end{eqnarray}
Minimization of Eq.~\eqref{Fen1} with respect to $\beta$ gives for the saturation field
\begin{equation}\label{FSfield1}
  h^{FAN}_s = S (J_\mathbf{k} - J_\mathbf{0} +2 E - 2 D)
\end{equation}
which is larger than its counterpart \eqref{ConeHs} for the conical phase by the quantity $S E$ reflecting the stabilization of the fan phase between the cone and the saturated states.

Terms containing $S^\eta_{\pm \mathbf{k}}$, $S^y_\mathbf{q} S^y_{-\mathbf{q}}$, and $S^z_\mathbf{q} S^z_{-\mathbf{q}}$ contribute to the linear part of the Hamiltonian $\mathcal{H}_1$ which contains only terms $a_{\pm (2n+1) \mathbf{k}}$ with factors of the order of $\beta^{2n+1}$, where $n$ is integer. In particular, we obtain for the factor before $a_\mathbf{k} + a_{-\mathbf{k}} + a^\dagger_\mathbf{k} + a^\dagger_{-\mathbf{k}} $ (which is the greatest term in $\mathcal{H}_1$)
\begin{eqnarray}
\label{FLin1}
  &&-\sqrt{\frac{NS}{2}} \frac{\beta}{2} \Bigl[ h - \left(\gamma -\frac{\varkappa}{2}\right) S (J_\mathbf{k}+2E) \\
  && + \gamma S (J_\mathbf{0} + 2D) -\frac{\varkappa}{2} S (J_{2\mathbf{k}} + 2D) + O(\beta^4) \Bigr]. \nonumber
\end{eqnarray}
Parameter $\beta$ should be chosen to make zero Eq.~\eqref{FLin1}. At $h< h^{FAN}_s$, one has
\begin{equation}
\label{Fbeta1}
  \beta^2 = \frac{8(h^{FAN}_s-h)}{S\left[ 3 J_\mathbf{k} - 2 J_\mathbf{0} - J_{2\mathbf{k}} + 6(E-D) \right]}.
\end{equation}
Eq.~\eqref{Fbeta1} minimizes also energy~\eqref{Fen1} that acquires the form (cf. Ref.~\cite{Nagamiya1962} for isotropic Heisenberg model)
\begin{equation}\label{Fen2}
  \varepsilon^{FAN}(h) = - h S - \frac{S^2}{2}(J_\mathbf{0} + 2 D) - \frac{S(h^{FAN}_s-h)^2}{3 h^{FAN}_s + S(J_\mathbf{0} - J_{2\mathbf{k}})}.
\end{equation}
Simple but tedious calculation shows that taking into account other terms in $\mathcal{H}_1$ provides corrections of the order of $\beta^6$ to Eq.~\eqref{Fen2}.

We proceed with analysis of the competition between the conical and the fan states. Results of Sec.~\ref{SCone} demonstrate that the transition between them can happen at $h^{FAN}_s - h \sim S E$ in which case $\beta^2 \sim S E$ (see Eq.~\eqref{Fbeta1}). The fan phase is stable if its spectrum near minimum (i.e., at $\mathbf{q} \approx \pm \mathbf{k}$) is positive that requires a positively defined bilinear part of the Hamiltonian $\mathcal{H}_2$ at $\mathbf{q} \approx \pm \mathbf{k}$ that reads as~\footnote{There are other relevant bilinear terms coupled via the umklapps to these ones, e.g., corresponding to momenta $\pm 3\mathbf{k}$. However, the latter come along with $S(J_\mathbf{k}-J_{3\mathbf{k}}) a^\dagger_{\pm 3\mathbf{k}} a_{\pm 3\mathbf{k}}$, which is evidently large positively defined contribution, thus not important for the spectrum stability. }
\begin{eqnarray}
\label{Fham3}
  \mathcal{H}_2' &=& C_\mathbf{k} (a^\dagger_\mathbf{k} a_\mathbf{k} + a^\dagger_{-\mathbf{k}} a_{-\mathbf{k}}) + B_\mathbf{k} (a^\dagger_\mathbf{k} a^\dagger _{-\mathbf{k}} + a_{\mathbf{k}} a_{-\mathbf{k}}) \nonumber \\
  && + U_\mathbf{k} (a^\dagger_\mathbf{k} a_{-\mathbf{k}} + a^\dagger_{-\mathbf{k}} a_{\mathbf{k}}) \\ &&+ V_\mathbf{k} (a^\dagger_\mathbf{k} a^\dagger_\mathbf{k} + a^\dagger_{-\mathbf{k}} a^\dagger_{-\mathbf{k}} + a_{\mathbf{k}} a_{\mathbf{k}} +  a_{-\mathbf{k}} a_{-\mathbf{k}}). \nonumber
\end{eqnarray}
Here
\begin{eqnarray}
  C_\mathbf{k} &=& \frac{S}{2} \left[ 2 E + \frac{\beta^2}{4} \left( J_\mathbf{k} - J_\mathbf{0} \right) \right], \nonumber\\
  B_\mathbf{k} &=& \frac{S}{2} \left[\frac{\beta^2}{4} \left( 2 J_\mathbf{k} - J_\mathbf{0} - J_{2\mathbf{k}} \right) - 2E \right], \\
  U_\mathbf{k} &=& \frac{S}{2} \frac{\beta^2}{4} \left( 2 J_\mathbf{k} - J_\mathbf{0} - J_{2\mathbf{k}} \right), \nonumber \\
  V_\mathbf{k} &=& \frac{S}{2} \frac{\beta^2}{8} \left( J_\mathbf{k} - J_\mathbf{0} \right). \nonumber
\end{eqnarray}
Eq.~\eqref{Fham3} can be equivalently written as
\begin{equation}\label{Fham4}
  \mathcal{H}_2' = \frac{1}{2} \psi^\dagger \hat{M} \psi,
\end{equation}
where we introduce vector
$
  \psi^\dagger = (a_\mathbf{k}, a_{-\mathbf{k}}, a^\dagger_\mathbf{k}, a^\dagger_{-\mathbf{k}}),
$
and the matrix
\begin{equation}\label{FM1}
  \hat{M} = \left(
              \begin{array}{cccc}
                C_\mathbf{k} & U_\mathbf{k} & 2 V_\mathbf{k} & B_\mathbf{k} \\
                U_\mathbf{k} & C_\mathbf{k} & B_\mathbf{k} & 2 V_\mathbf{k} \\
                2 V_\mathbf{k} & B_\mathbf{k} & C_\mathbf{k} & U_\mathbf{k} \\
                B_\mathbf{k} & 2 V_\mathbf{k} & U_\mathbf{k} & C_\mathbf{k} \\
              \end{array}
            \right).
\end{equation}
The fan phase is stable if this matrix is positively defined. Equivalently, all its eigenvalues should be positive or at least zero. Details on eigenvalues calculation are presented in Appendix~\ref{AppendixB}, where we show that one of eigenvalues having the form
\begin{equation}\label{Fev1}
  C_\mathbf{k} - B_\mathbf{k} - U_\mathbf{k} +2 V_\mathbf{k} = \frac{S}{2} \left[ 4 E - \frac{\beta^2}{2}(J_\mathbf{k}- J_{2 \mathbf{k}})\right]
\end{equation}
becomes negative upon $\beta$ growth (i.e., upon the field decreasing) at
\begin{equation}\label{Fbcr}
  \beta^2>\beta^2_{cr} = \frac{8 E}{J_\mathbf{k} - J_{2 \mathbf{k}}}.
\end{equation}
It follows from Eqs.~\eqref{Fbeta1} and \eqref{Fbcr} that the fan phase is stable at $h>h_{cr}$, where
\begin{eqnarray}
\label{FHcr1}
  h_{cr} &=& h^{FAN}_s - S E \frac{3J_\mathbf{k} - 2 J_ \mathbf{0}- J_{2 \mathbf{k}}}{J_\mathbf{k} - J_{2 \mathbf{k}}} \\ \nonumber &=& h^{XY}_s - 2 S E \frac{J_\mathbf{k}- J_{\mathbf{0}}}{J_\mathbf{k}- J_{2\mathbf{k}}}
\end{eqnarray}
in the leading order in $E/J$. Our numerics justify this result (see Appendix~\ref{AppendixC}). Notice that this accurate derivation of $h_{cr}$ provides only the additional factor of $2$ in the second term in comparison with the simple estimation \eqref{HCr}.

It should be noted that we assume $D\agt E$ in the present section and in Sec.~\ref{SCone}. However an accurate consideration shows that one can treat constant $D$ not as a small quantity. The only restriction on $D$ is that it should not be large enough to destroy the conical and the fan phases. The modification of the results obtained above is very simple for arbitrary $D$: each $J_\mathbf{0}$ and $J_{2\mathbf{k}}$ should come along with term $2 D$. For example, the counterpart of Eq.~\eqref{FHcr1} reads as
\begin{eqnarray}\label{FHcr2}
  h_{cr} &=& h^{FAN}_s - S E \frac{3J_\mathbf{k} - 2 J_ \mathbf{0}- J_{2 \mathbf{k}} - 6 D}{J_\mathbf{k} - J_{2 \mathbf{k}} -2 D}.
\end{eqnarray}

Finally, let us discuss other orientations of magnetic field: along the medium and the hard axes. The results obtained above are modified simply in these cases. If the field is applied along the medium $y$ axis, one can simply interchange $E$ with $D$ provided that $D \ll J$. For the field directed along the hard $x$ axis, the effective anisotropy in the spiral plane is $D-E$. Then, one has to substitute $E$ with $D-E$ and $D$ with $-E$. In both cases, the spiral plane remains anisotropic and the transition to the fan phase occurs.

\section{Possible sequences of phase transitions near the saturation field}
\label{SSeq}

In this section, we discuss possible sequences of phase transitions when the anisotropy is moderate so that some kind of commensurate antiferromagnetic ordering characterized by momentum $\mathbf{k}_0$ can compete with the helical structure. The analysis below is similar to that of Ref.~\cite{utesov2019} performed at small field. Importantly, phase transitions involving two-up-two-down $\uparrow \uparrow \downarrow \downarrow$ structure can be discussed in the same way (see also Sec.~\ref{SComp}).

At moderate anisotropy, one should take into account also the canted antiferromagnetic phase (CAF) whose classical energy per spin reads as
\begin{eqnarray}\label{ECaf1}
  \varepsilon^{CAF} &=& -\frac{S^2 \sin^2{\alpha}}{2}  ( J_\mathbf{0} - J_{\mathbf{k}_0} + 2D - 2E) \nonumber \\
  && -\frac{S^2}{2} (J_{\mathbf{k}_0} + 2E)  - h S \sin{\alpha},
\end{eqnarray}
where we assume that $\mathbf{h}$ is directed along $z$ axis and $\alpha$ is the canting angle. Minimization of Eq.~\eqref{ECaf1} gives
\begin{equation}\label{CafAl}
  \sin{\alpha} = \frac{h}{h^{CAF}_s},
\end{equation}
where
\begin{equation}\label{CafSat}
  h^{CAF}_s = S(J_{\mathbf{k}_0} -  J_\mathbf{0} - 2D + 2E).
\end{equation}
Substitution of Eq.~\eqref{CafAl} into Eq.~\eqref{ECaf1} gives
\begin{eqnarray}\label{ECaf2}
  \varepsilon^{CAF}(h) &=& -\frac{S^2}{2} (J_{\mathbf{k}_0} + 2E)  - \frac{S h^2}{2 h^{CAF}_s}.
\end{eqnarray}
The corresponding formula for the conical phase without anisotropy-induced corrections reads as
\begin{equation}\label{exy2}
  \varepsilon^{XY}_0(h) = -\frac{S^2}{2} (J_{\mathbf{k}} + E)  - \frac{S h^2}{2 h^{XY}_s}.
\end{equation}
It is easy to show that
\begin{equation}
\label{CAFXY1}
  \varepsilon^{CAF}(h) - \varepsilon^{XY}_0(h) = \frac{S^2(F-E)}{2} \left( 1 -\frac{h^2}{h^{CAF}_s h^{XY}_s} \right),
\end{equation}
where $F = J_{\mathbf{k}} - J_{\mathbf{k}_0} $.
The expression in the last brackets of Eq.~\eqref{CAFXY1} is positive and the overall sign of the energy difference is determined by $F-E$. If the latter is positive, XY phase has lower energy and the consideration performed in previous sections remains valid. In the opposite case of $E - F >0$ the CAF phase comes into play at moderate $h\lesssim h_{cr}$ and, as we show below, the first order transition to either conical or the fan phase occurs.

Let us discuss first the possibility of CAF$\leftrightarrow$XY transition. It appears due to the correction~\eqref{EnCorrXY1} to the classical energy of XY state (we use below simplified expression \eqref{EnCorrXY2} for this correction). Then, the transition field $h_1$ satisfies the equation
\begin{equation}\label{CAFXY2}
  \frac{E -F}{2} \left( 1 -\frac{h^2_1}{h^{CAF}_s h^{XY}_s} \right) = \frac{E^2}{J_\mathbf{k} - J_{2\mathbf{k}}}
\end{equation}
which gives after some algebra
\begin{equation}\label{CAFXY3}
  h_1 = h^{XY}_s - S \left[ \frac{E^2}{E-F} \frac{J_\mathbf{k} - J_{\mathbf{0}}}{J_\mathbf{k} - J_{2\mathbf{k}}} - \frac{E-F}{2} \right].
\end{equation}
This field should be lower than $h_{cr}$ given by Eq.~\eqref{FHcr1} in order the transition CAF$\leftrightarrow$XY can occur. We find after tedious calculations that this condition is equivalent to
\begin{equation}
\label{CAFXY4}
  E < F \left( 1 + \sqrt{\frac{2 (J_\mathbf{k} - J_{\mathbf{0}})}{ 3 J_\mathbf{k} - 2 J_{\mathbf{0}} - J_{2\mathbf{k}}  } }\right).
\end{equation}
Thus, one has CAF$\leftrightarrow$XY$\leftrightarrow$FAN sequence of phase transitions if Eq.~\eqref{CAFXY4} holds.

Let us discuss the possibility of CAF$\leftrightarrow$FAN transition.
One obtains from Eqs.~\eqref{Fen2} and \eqref{ECaf2}
\begin{equation}\label{CAFFAN1}
  \varepsilon^{CAF}(\delta h) - \varepsilon^{FAN}(\delta h) = \frac{(S F + \delta h)^2}{3 J_\mathbf{k} - 2 J_{\mathbf{0}} - J_{2\mathbf{k}}} - \frac{S \delta h^2}{2 h^{CAF}_s},
\end{equation}
where $\delta h = h^{CAF}_s - h \geq 0$ and we use that $h^{FAN}_s = h^{CAF}_s + S F$.
It is clear from Eq.~\eqref{CAFFAN1} that the fan phase is always energetically preferable in the considered here case of $F > 0 $ at $\delta h \ll F$. At the transition point $h_2$, the right-hand side of Eq.~\eqref{CAFFAN1} is zero and we obtain after some algebra
\begin{equation}\label{CAFFAN2}
  h_2 = h^{FAN}_s - \frac{S F}{\displaystyle{1 - \sqrt{\frac{2 (J_\mathbf{k} - J_{\mathbf{0}})}{ 3 J_\mathbf{k} - 2 J_{\mathbf{0}} - J_{2\mathbf{k}}}}}}.
\end{equation}
Notice that the denominator in the second term in Eq.~\eqref{CAFFAN2} is always positive because
\begin{equation}
  \frac{2 (J_\mathbf{k} - J_{\mathbf{0}})}{ 3 J_\mathbf{k} - 2 J_{\mathbf{0}} - J_{2\mathbf{k}}} = \frac{\frac{2 (J_\mathbf{k} - J_{\mathbf{0}})}{ J_\mathbf{k} - J_{2\mathbf{k}} }}{1 + \frac{2 (J_\mathbf{k} - J_{\mathbf{0}})}{ J_\mathbf{k} - J_{2\mathbf{k}}}} < 1.
\end{equation}
Another restriction on $h_2$ is that it should be larger than $h_{cr}$. As a result, we obtain that CAF$\leftrightarrow$FAN transition takes place when
\begin{equation}
\label{CAFFAN3}
  E > F \left( 1 + \sqrt{\frac{2 (J_\mathbf{k} - J_{\mathbf{0}})}{ 3 J_\mathbf{k} - 2 J_{\mathbf{0}} - J_{2\mathbf{k}}  } }\right)
\end{equation}
which is complementary to the condition \eqref{CAFXY4} of CAF$\leftrightarrow$XY$\leftrightarrow$FAN transition.

The final remark in the previous section on the $D$ value is applicable also for the results of the present section: one can introduce $D$ to Eqs.~\eqref{CAFXY3} and \eqref{CAFFAN2}, and consider the field direction along the medium or the hard axes.

\section{Comparison with previous works}
\label{SComp}

We compare predictions of our analytical approach with results of Ref.~\cite{zh} in which an anisotropic next-nearest neighbor Heisenberg (ANNNH) model was considered by means of a real-space mean-field approach. The complicated phase diagram on the temperature-field plane of multiferroic MnWO$_4$ was successfully reproduced in Ref.~\cite{zh} (see, also Refs.~\cite{Ehrenberg1997,mnw3,mnw2,Quirion2013}). In ANNNH model discussed in Ref.~\cite{zh}, spins interact ferromagnetically within $ab$ plane and there is a frustrating antiferromagnetic interaction along $c$ axis with antiferromagnetic couplings $J_1=-1$ and $J_2 = - 2$ between nearest and next-nearest neighbors, respectively, so that
\begin{equation}
\label{Mn1}
  J_\mathbf{q} = 2 (J_1 \cos{q_c} + J_2 \cos{2 q_c}).
\end{equation}
On the mean-field level, this model is equivalent to a classical spin chain along $c$ axis because each ferromagnetic $ab$ plane plays the role of a classical spin. Then, $\mathbf{k} = (0,0, 0.54 \pi)$ and the competing spin structure is two-up-two-down $\uparrow \uparrow \downarrow \downarrow$ one with $\mathbf{k}_0 =(0,0, \pi/2)$ that yields $F = J_{\mathbf{k}}-J_{\mathbf{k}_0} \approx 0.125$. Other parameters used in Ref.~\cite{zh} were
\begin{equation}
\label{Mn2}
  D=0.4, \quad E=0.2, \quad S=5/2.
\end{equation}
According to our consideration above, when magnetic field is directed along the easy or the medium axes, CAF phase is the ground state at moderate fields that was indeed observed in Ref.~\cite{zh}.

For the field directed along $z$ axis, our analytical approach gives $h_1 \approx 21$, $h_{cr} \approx 21.9$, and $h^{FAN}_s \approx 24.3$ for the critical fields in the sequence CAF$\leftrightarrow$XY$\leftrightarrow$FAN of phase transitions. These values are in quite good agreement with the numerical results of Ref.~\cite{zh} $h_1 \approx 20.25$, $h_{cr} \approx 22.4$ and $h^{FAN}_s \approx 24.4$ despite rather large values of anisotropy constants. Together with our findings of Ref.~\cite{utesov2019}, the present theory provides an analytical description of the sequence of five magnetic field-induced phase transitions at small temperatures in this model observed also experimentally in MnWO$_4$.

If the field is oriented along the medium $y$ axis, $D=0.4$ in Eq.~\eqref{Mn2} brings the model out of the formal domain of validity of our theory. In particular, one obtains from Eq.~\eqref{Fbcr} $\beta_{cr} \approx 0.78$ so that there is no required small parameter for the fan phase consideration. Nevertheless, our results work good even in this case. Because $h_1>h_{cr}$, there is CAF$\leftrightarrow$FAN transition. We get $h_2 \approx 23.55$ from Eq.~\eqref{CAFFAN2} that is in good agreement with the value $h_2 \approx 23$ observed in Ref.~\cite{zh} because the difference is much smaller than $h^{FAN}_s - h_2$. One has $h^{FAN}_s \approx 26.3$ from Eq.~\eqref{FSfield1} which is in a good agreement with the numerically obtained value $h^{FAN}_s \approx 26.1$.


\section{Dipolar forces}
\label{SDip}

One of the possible sources of the biaxial anisotropy is the dipolar interaction~\cite{Utesov2017}. It is of prime importance when magnetic ions are in $L=0$ state with half-filled electronic shell as in Mn$^{2+}$ or Eu$^{2+}$ because the spin-orbit interaction is particularly small in this case. We consider in this section Hamiltonian~\eqref{ham1} in which $\mathcal{H}_{an}$ is replaced by
\begin{equation}
 \label{Hdip1}
  \mathcal{H}_d = \frac12 \sum_{i,j} D^{\alpha \beta}_{ij} S^\alpha_i S^\beta_j,
\end{equation}
where
\begin{equation}\label{dip1}
	 {\cal D}^{\alpha \beta}_{ij} = \omega_0 \frac{v_0}{4 \pi} \left( \frac{1}{R_{ij}^3} - \frac{3 R_{ij}^\alpha R_{ij}^\beta }{R_{ij}^5}\right),
\end{equation}
$v_0$ is a unit cell volume, and
\begin{equation}\label{dipen}
  \omega_0 = 4 \pi \frac{(g \mu_B)^2}{v_0} \ll J
\end{equation}
is the characteristic energy of the dipole interaction.

After Fourier transform~\eqref{four1} one obtains
\begin{equation}\label{dip2}
  \mathcal{H}_d = \frac12 \sum_\mathbf{q} {\cal D}^{\alpha \beta}_\mathbf{q} S^\alpha_\mathbf{q} S^\beta_{-\mathbf{q}}.
\end{equation}
At $\mathbf{q} = 0$, tensor ${\cal D}^{\alpha \beta}_\mathbf{0}$ should be substituted by $\omega_0 \mathcal{N}^{\alpha \beta}$, where $\mathcal{N}^{\alpha \beta}$ is the demagnetization tensor~\cite{SpinWaves} and we assume the shape of the sample to be an ellipsoid. For each momentum $\mathbf{q}$, tensor ${\cal D}^{\alpha \beta}_\mathbf{q}/2$ has a set of eigenvalues $\lambda_1(\mathbf{q}) \geq \lambda_2(\mathbf{q}) \geq \lambda_3(\mathbf{q})$ and corresponding orthogonal eigenvectors $\mathbf{v}_1(\mathbf{q})$, $\mathbf{v}_2(\mathbf{q})$, and $\mathbf{v}_3(\mathbf{q})$. The latter determine the hard, the medium and the easy axes for each particular momentum. As before, we can neglect a small influence of the anisotropy on the magnetic structure vector $\bf k$ in the conical and in the fan phases.

Results found above become valid in this case after some substitutions. Let us direct $z$, $y$, and $x$ axes along $\mathbf{v}_3(\mathbf{q})$, $\mathbf{v}_2(\mathbf{q})$, and $\mathbf{v}_1(\mathbf{q})$, respectively. When the field is parallel to $z$, the role of $E$ is played by $\lambda_1(\mathbf{k}) - \lambda_2(\mathbf{k})$.
$D = \lambda_1(\mathbf{k}) - \omega_0 \mathcal{N}_{zz}/2$, where the last term defines a positive contribution to the system energy from the demagnetization field on which value ($4 \pi \mathcal{N}_{zz} M$, where $M$ is the magnetization in the saturated phase) the saturation field \eqref{FSfield1} increases. In a simple case of the sample in the form of a cylinder with the axis parallel to $\mathbf{v}_3(\mathbf{k})$, one has $N_{zz}=0$, $E=\lambda_1(\mathbf{k}) - \lambda_2(\mathbf{k})$, and $D = \lambda_1(\mathbf{k})$. Then, dipolar forces can be responsible for XY$\leftrightarrow$FAN transition at $\lambda_1(\mathbf{k}) \neq \lambda_2(\mathbf{k})$.

Other field directions can be analyzed similarly. For the field directed along $y$ axis, $E = \lambda_1(\mathbf{k}) - \lambda_3(\mathbf{k})$ and $D = \lambda_1(\mathbf{k}) - \omega_0 \mathcal{N}_{yy}/2$. If $\mathbf{h}$ is parallel to $x$ axis, $E = \lambda_2(\mathbf{k}) - \lambda_3(\mathbf{k})$ and $D = \lambda_2(\mathbf{k}) - \omega_0 \mathcal{N}_{xx}/2$.

Notice also that if $\lambda_i(\mathbf{k}_0) \approx \lambda_i(\mathbf{k})$, one can use results of Sec.~\ref{SSeq} for phase transitions involving CAF phase.

\section{Summary and conclusion}
\label{SSum}

To summarize, we demonstrate that small biaxial anisotropy and/or dipolar interaction can lead to a subtle competition between canted antiferromagnetic, cone helical, and fan states in frustrated antiferromagnet at strong magnetic field near its saturation value. As a result, different sequences of phase transitions can appear depending on values of anisotropic interactions which are summarized in Fig.~\ref{fig1}. We perform an analytical mean-field consideration and derive ground-state energies of all spin states and critical fields of all transitions which are in a very good quantitative agreement with our and previous Monte-Carlo simulations. The present strong-field consideration together with our previous discussion \cite{Ueda2009} of the weak-field regime provides, in particular, an analytical mean-field description of all field-induced phase transitions observed experimentally in multiferroic MnWO$_4$.

\begin{acknowledgments}

The reported study was funded by RFBR according to the research project 18-02-00706.

\end{acknowledgments}

\appendix

\section{Magnon spectrum in the conical phase}
\label{AppendixA}

Here we discuss the classical magnon spectrum in the conical phase
which can be obtained using Eqs.~\eqref{hamB1}--\eqref{bq} bearing in mind that $C_\mathbf{q} \neq C_{-\mathbf{q}}$, the result being
\begin{equation}\label{SpecM1}
  \epsilon_\mathbf{q} = \frac{C_\mathbf{q} - C_{-\mathbf{q}}}{2} + \sqrt{\left( \frac{C_\mathbf{q} + C_{-\mathbf{q}}}{2} \right)^2 - B^2_\mathbf{q}}.
\end{equation}
Notice that $\epsilon_\mathbf{0} = 0$ at any field.

\begin{figure}[b]
  \centering
  \includegraphics[width=0.9\linewidth]{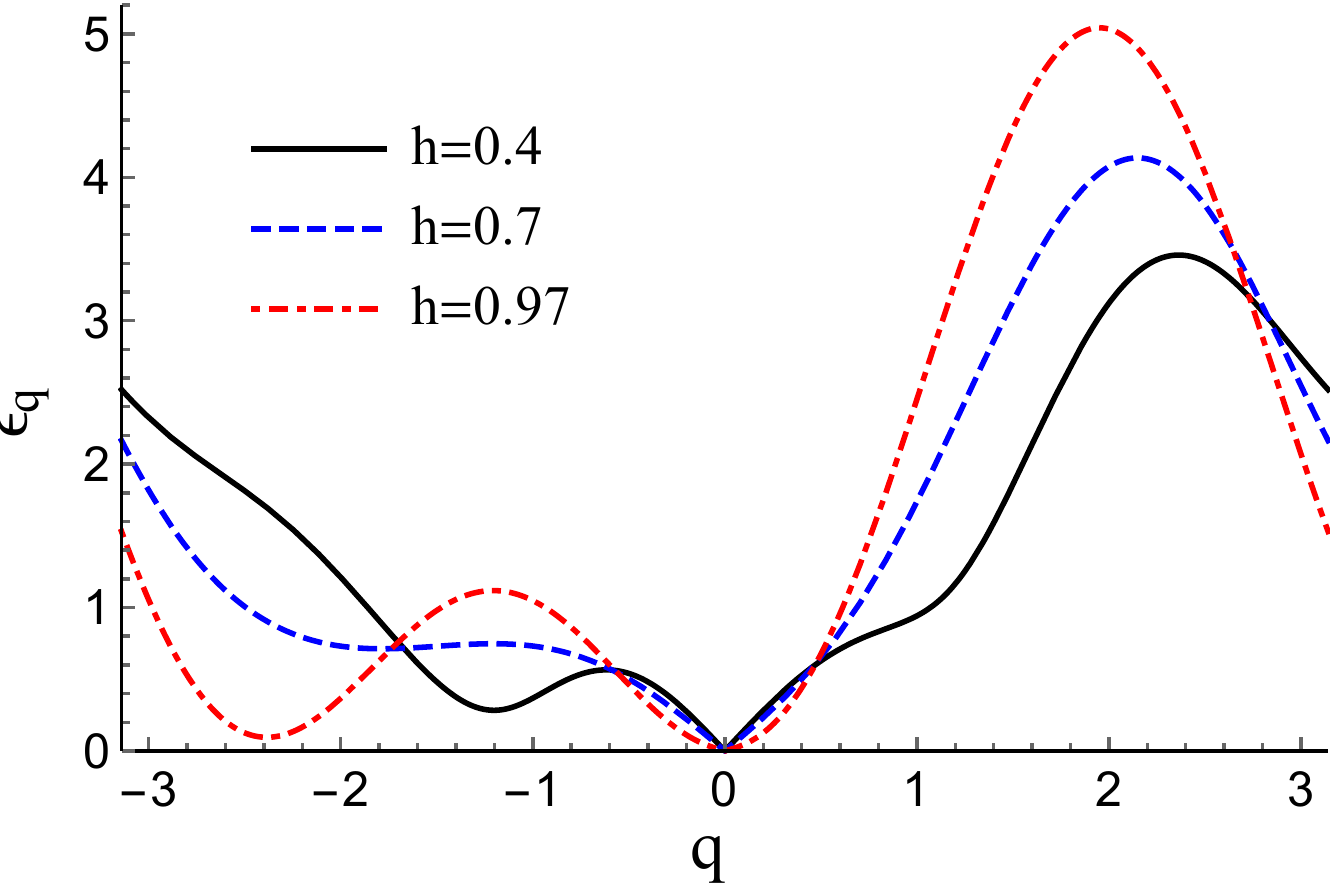}\\
  \caption{Magnon spectra for ANNNH model in the conical phase at different magnetic fields (see the text). Importantly, spectrum is essentially asymmetric with respect to  the momentum inversion. Notice a deep roton-like minimum at ${\bf q}=-2 \mathbf{k}$ at fields close to the critical one $h_{cr} \approx 0.977$.
	\label{figA}}
\end{figure}

We draw spectrum \eqref{SpecM1} in Fig.~\ref{figA} for $h=0.4$, 0.7, and 0.97 directed along $z$ axis for ANNNH model described in Sec.~\ref{SComp} with $J_1=1$, $J_2=-0.7$, $D=0.1$, $E=0.05$, and $S=1$. For these parameters, $\mathbf{k} \approx 0.38 \pi$, the spiral plane flop field~\cite{utesov2018} is $h_{sp} \approx 0.394$, $h^{FAN}_s \approx 1.057$, and $h_{cr} \approx 0.977$. One can see in Fig.~\ref{figA} spectra asymmetry according to the momentum inversion. At small fields, there are some features at ${\bf q}=\pm \mathbf{k}$ (the spectrum is zero at these momenta at $E=D=h=0$). Then, at large field close to $h_{cr}$, the spectrum has a roton-like minimum at ${\bf q}=-2\mathbf{k}$ which arises by the following reason. Notice that
\begin{equation}\label{AppA1}
  C_\mathbf{q} = S \left( J_\mathbf{k} - J_{\mathbf{k}+\mathbf{q}} \right) + O(\tilde{\alpha}^2),
\end{equation}
where $\tilde{\alpha}=\pi/2 - \alpha$ (see Sec.~\ref{SCone}) and the first term is exactly zero at $\mathbf{q}=\mathbf{0},-2\mathbf{k}$ so $C_\mathbf{q}, B_\mathbf{q} \sim \tilde\alpha^2$. Then, it is easy to show that
\begin{equation}\label{AppA2}
  \epsilon_{-2\mathbf{k}} \approx \frac{S}{2} \tilde{\alpha}^2 (J_\mathbf{k}-J_{2\mathbf{k}})
\end{equation}
which is a small quantity near the transition to the fan phase. We speculate that this roton-like minimum touches zero at the critical field $h_{cr}$ and the condensation of magnons with ${\bf q}=-2\mathbf{k}$ at larger magnetic fields restores the broken $\mathbb{Z}_2$ symmetry.

\section{Eigenvalues of $\hat{M}$}
\label{AppendixB}

Here we derive eigenvalues of matrix~\eqref{FM1}. Using Pauli matrices $\sigma_{0,x}$, we can write the characteristic equation in the form
\begin{equation}\label{AppB1}
  \left|
    \begin{array}{cc}
      (C_\mathbf{k} - \lambda) \sigma_0 + U_\mathbf{k} \sigma_x & 2 V_\mathbf{k} \sigma_0 + B_\mathbf{k} \sigma_x \\
      2 V_\mathbf{k} \sigma_0 + B_\mathbf{k} \sigma_x & (C_\mathbf{k} - \lambda) \sigma_0 + U_\mathbf{k} \sigma_x \\
    \end{array}
  \right| =0
\end{equation}
which is equivalent to
\begin{eqnarray}\label{AppB2}
  \bigl| \left[ (C_\mathbf{k} - \lambda)^2 + U^2_\mathbf{k} - 4 V^2_\mathbf{k} - B^2_\mathbf{k}  \right]\sigma_0- \\ 2 \left[ (C_\mathbf{k} - \lambda) U_\mathbf{k} -  2 V_\mathbf{k} B_\mathbf{k}\right] \sigma_x \bigr| =0 \nonumber
\end{eqnarray}
and to
\begin{eqnarray}\label{AppB3}
  \left[ (C_\mathbf{k} - \lambda)^2 + U^2_\mathbf{k} - 4 V^2_\mathbf{k} - B^2_\mathbf{k}  \right]^2 \\ -4 \left[ (C_\mathbf{k} - \lambda) U_\mathbf{k} -  2 V_\mathbf{k} B_\mathbf{k}\right]^2=0.  \nonumber
\end{eqnarray}
Solutions of Eq.~\eqref{AppB3} have the form
\begin{eqnarray}\label{AppB4}
  \lambda &=& C_\mathbf{k} + U_\mathbf{k} \pm (B_\mathbf{k} + 2 V_\mathbf{k}), \nonumber\\
	\lambda &=& C_\mathbf{k} - U_\mathbf{k} \pm (B_\mathbf{k} - 2 V_\mathbf{k}).
\end{eqnarray}

\section{Numerics}
\label{AppendixC}

Here we present some details of our numerical modeling. We utilize the Monte-Carlo method in order to find classical energy of different phases in ANNNH model (see Sec.~\ref{SComp}) at zero temperature. We perform $10^6 - 10^7$ steps considering chains with 1000 and 2000 sites. Our numerical findings strongly support analytical results for the conical phase energy including anisotropy-induced correction \eqref{EnCorrXY2}, the fan phase energy \eqref{Fen2}, and the critical fields \eqref{FHcr1}, \eqref{CAFXY3}, and \eqref{CAFFAN2}.

As a particular example, we present here results for parameters $J_1=1$, $J_2=-0.7$, $D=0.1$, $E=0.05$, and $S=1$ (as in Appendix~\ref{AppendixA}). The scenario shown in Fig.~\ref{fig1}(b) takes place in this case. We quantify the field of the phase transition XY$\leftrightarrow$FAN as a point at which the average $(S^x_j)^2$ becomes smaller than $10^{-6}$.
The numerical result $h_{cr} \approx 0.976$ is very close to the analytical finding $h_{cr}\approx 0.977$ obtained from Eq.~\eqref{FHcr1}. At $h=0.97$, the energy per spin in the conical phase $\approx 1.3714$ coincides with the analytical prediction with correction \eqref{EnCorrXY2} up to the fourth decimal digit. For the fan phase at $h=0.98$, numerically obtained energy per spin is $\approx 1.3811$ whereas Eq.~\eqref{Fen2} gives $\approx 1.3812$. We observe numerically and analytically for $\beta$ $0.335$ and $0.329$, respectively.

\bibliography{TAFbib}

\end{document}